\documentclass[sigconf]{acmart}

\usepackage{booktabs} 
\usepackage{subfigure}
\usepackage{multirow}
\usepackage{array}
\usepackage{setspace}
\usepackage[skip=0pt]{caption}
\usepackage{enumitem}

\copyrightyear{2018} 
\acmYear{2018} 
\setcopyright{acmlicensed}
\acmConference[WSDM 2018]{WSDM 2018: The Eleventh ACM International Conference on Web Search and Data Mining }{February 5--9, 2018}{Marina Del Rey, CA, USA}
\acmPrice{15.00}
\acmDOI{10.1145/3159652.3159686}
\acmISBN{978-1-4503-5581-0/18/02}

\newcommand{\negskip}{\vspace*{-0.5\baselineskip}}

\begin{document}
\begin{spacing}{0.96}

\title{Why People Search for Images using Web Search Engines}

\author{Xiaohui Xie}
\orcid{}
\affiliation{%
\institution{DCST, Tsinghua University}
\city{Beijing}
\country{China}
}
\email{xiexh_thu@163.com}

\author{Yiqun Liu}
\orcid{}
\affiliation{%
\institution{DCST, Tsinghua University}
\city{Beijing}
\country{China}
}
\email{yiqunliu@tsinghua.edu.cn}

\author{Maarten de Rijke}
\orcid{0000-0002-1086-0202}
\affiliation{%
\institution{University of Amsterdam}
\city{Amsterdam}
\country{The Netherlands}
}
\email{derijke@uva.nl}

\author{Jiyin He}
\orcid{}
\affiliation{%
\institution{CWI}
\city{Amsterdam}
\country{The Netherlands}
}
\email{j.he@cwi.nl}

\author{Min Zhang}
\orcid{}
\affiliation{%
\institution{DCST, Tsinghua University}
\city{Beijing}
\country{China}
}
\email{z-m@tsinghua.edu.cn}

\author{Shaoping Ma}
\orcid{}
\affiliation{%
\institution{DCST, Tsinghua University}
\city{Beijing}
\country{China}
}
\email{msp@tsinghua.edu.cn}

\renewcommand{\shortauthors}{X. Xie et al.}

\begin{abstract}
What are the intents or goals behind human interactions with image search engines? Knowing why people search for images is of major concern to Web image search engines because user satisfaction may vary as intent varies. Previous analyses of image search behavior have mostly been query-based, focusing on what images people search for, rather than intent-based, that is, why people search for images. To date, there is no thorough investigation of how different image search intents affect users' search behavior.
    
In this paper, we address the following questions: (1)~Why do people search for images in text-based Web image search systems? (2)~How does image search behavior change with user intent? (3)~Can we predict user intent effectively from interactions during the early stages of a search session? To this end, we conduct both a lab-based user study and a commercial search log analysis.
    
We show that user intents in image search can be grouped into three classes: Explore/Learn, Entertain, and Locate/Acquire. Our lab-based user study reveals different user behavior patterns under these three intents, such as first click time, query reformulation, dwell time and mouse movement on the result page. Based on user interaction features during the early stages of an image search session, that is, before mouse scroll, we develop an intent classifier that is able to achieve promising results for classifying intents into our three intent classes. Given that all features can be obtained online and unobtrusively, the predicted intents can provide guidance for choosing ranking methods immediately after scrolling.

\end{abstract}

%
%



\keywords{Image search; User intent; User behavior}

\maketitle

\section{Introduction}\label{section:introduction}
Intent is assumed to be the immediate antecedent of behavior~\cite{aizen2010behavioral,huurnink2010search}. For any information access service, it is important to understand the underlying intent behind user behavior. A widely used Web search intent taxonomy was proposed by~\citet{broder2002taxonomy}, based on log-based user studies. In Broder's taxonomy, search intent of Web search users is categorized into three classes: informational, transactional and navigational.

The diversification of online information and online information services that we have seen since the introduction of Broder's search intent taxonomy creates challenges to this taxonomy. Hence, several refinements of this taxonomy have been proposed~\citep{rose2004understanding,russell2009task}. However, few efforts have been made towards understanding the intents of Web image search users. In image search engines, the items users search for are images instead of Web pages or online services. And the different result placement and interaction mechanisms of image search also make the search process rather different from general Web search engines~\citep{xie2017investigating}. \citet{kofler2009dynamic} supported this view by conducting a user study and showing that without adaptation, user intent taxonomies applied in general web search are not applicable for image search. Similar to general purpose Web search~\cite{hu2015search,dai2006detecting}, we believe that a thorough understanding or even successful detection of users' image search intent helps search engines provide better image search results and improve their satisfaction. This motivates our first research question:
\begin{itemize}[leftmargin=25pt]
\item[\emph{\textbf{RQ1:}}] \emph{\textbf{Why do people search for images in text-based Web image search systems?} }
\end{itemize}
User behavior data has been successfully adopted to improve general Web searches in result ranking~\cite{agichtein2006improving,xu2012incorporating}, query suggestion~\cite{cao2008context,wang2010inferring}, query auto completion~\cite{jiang2014learning,li2014two}, etc. We therefore believe that understanding user interaction behavior in the multimedia search scenarios will also provide valuable insight into the understanding of user intent. As user behavior varies with search intent, looking into differences in behavior and how such differences relate to search intent will help to improve the performance of image search engines.

Previous work on image search intent understanding usually focuses on the query proposed by users and assumes that the query represents user intent well. 
However, determining users' search intents or information needs based on the queries they submitted is sometimes rather difficult, as a large proportion of keyword based queries are short, ambiguous or broad~\cite{dou2007large,jansen2000real,rafiei2010diversifying}. Compared to general (Web) search, queries used in image search on the Web tend to be even shorter~\cite{goodrum1999visual}. As the same query may come from different search intents, previous work attempts to bridge the \textbf{intent gap} between query and the underlying intent through explicit methods, including query suggestion~\cite{zha2010visual,xu1996query} and result diversification~\cite{van2009visual,taneva2010gathering}. Besides work that investigates user intent based on query analysis, others look into the relationship between what people search for (query-based) and how they interact with image search engines. For example, \citet{park2015large} categorize queries using two orthogonal taxonomies (subject-based and facet-based) and identify important behavioral differences across query types. Unlike previous work, we avoid analyzing query content and, instead, provide a thorough investigation into the whole interaction processes of users. As user behavior is directly affected by search intent, we propose an intent taxonomy based on the differences in user behavior patterns. This motivates our second research question:
\begin{itemize}[leftmargin=25pt]
\item[\emph{\textbf{RQ2:}}]
\emph{\textbf{How does image search behavior change with user intent?}}
\end{itemize}
Automatically identifying search intent at an early stage of a search session helps a multimedia retrieval system to rerank its results according to the underlying user intent. Since pagination is usually not (explicitly) supported on image search SERPs, users can view results by scrolling up and down instead of clicking on the ``next page'' button. In this paper, we define the ``early stage'' of a search session as search behavior before any scrolling takes place. As user interactions with image search engines through mouse and keyboard can be captured online and unobtrusively, it will be practical to build intent recognition system based on these features if they are effective. This motivates our third research question:
\begin{itemize}[leftmargin=25pt]
\item[\emph{\textbf{RQ3:}}]
\emph{\textbf{Can we effectively predict user intent from user interactions at the early stage of a search session?}}
\end{itemize}
To summarize, the main contributions of this work are:
\begin{itemize}[nosep,leftmargin=10pt]
	\item We propose a new image search intent taxonomy based on an open-coded discussion methodology~\cite{russell2009task}. The taxonomy includes three classes: Explore/Learn, Entertain, Locate/Acquire. As far as we know, ours is the first work to focus on an image search intent taxonomy in Web search engines. 
	We verify the proposed image search intent taxonomy in two ways: through a user survey involving over 200 people and through a Web image search log analysis. Results show that the taxonomy covers a majority of user intents in Web image search.  
	\item We design 12 tasks within the scope of the taxonomy and perform a lab-based user study to show that users interact with image search engines in different ways with different information needs. Differences are observed in temporal patterns, query reformulation patterns and mouse movement patterns. 
	\item We build and evaluate a user intent recognition system based on the user interactions at the early stage of search sessions and achieve state-of-the-art prediction results.
\end{itemize}
%
The rest of the paper is organized as follows. Section~\ref{section:related_work} reviews related work. Section~\ref{section:userintenttaxonomy} presents our user intent taxonomy in image search. Section~\ref{section:imagesearchwithdifferentintents} introduces our user-study settings and reports the findings. Intent prediction and its results are given in Section~\ref{section:imagesearchintentprediction}. Finally, Section~\ref{section:conclusionandfuturework} discusses conclusions and future work.

\section{Related work}\label{section:related_work}

\subsection{Search intent taxonomies}\label{subsection:searchintenttaxomomies}
In the past two decades intent taxonomies for general Web search have been investigated by several researchers. \citet{broder2002taxonomy} introduced a taxonomy of user intent in text-based Web search using both randomly selected search queries and an analysis of survey data collected from AltaVista users. The taxonomy consists of three categories: Informational, Navigational and Transactional. In \cite{jansen2007query}, session characteristics of these three top-level search intents were examined and used to develop a classification algorithm. Building on Broder's taxonomy, \citet{rose2004understanding} introduced a sub-classification of the taxonomy to classify intent more precisely. By having humans assign task-type labels to search sessions based on Rose and Levinson's taxonomy, \citet{russell2009task} found that it is hard to get sufficient inter-rater agreement on ambiguous search tasks and proposed a new search task taxonomy that contains seven categories. Taxonomies investigated in general Web search mentioned above were found to be \emph{not} applicable in image search~\cite{kofler2009dynamic,kofler2009exploratory}.

Previous work on an intent taxonomy for image search mostly focused on \emph{what} people search for not \emph{why} they search. \citet{pu2005comparative} classified 1,000 frequent image queries based on a proprietary subject-based categorization scheme. By focusing on whether users were searching for people, location, etc., and on whether the search was about unique instances or non-unique instances, \citet{jansen2008searching} classified queries based on three non subject-based image query classification schemas. \citet{lux2010classification} are among the first to investigate the image search intent problem. They categorized user intents into knowledge orientation, mental image, navigation, and transaction; these intents describe search activities in Flickr, a digital photo sharing platform instead of a search engine. After that, a two-dimensional taxonomy was proposed in~\cite{cheng2017predicting}. However, this work was based on another sharing platform (Pinterest). Neither works were conducted in Web search engines which may face more complex search scenarios. \citet{redi2017bridging} showed that Web image search must deal with images from a wide variety of sources including very poor quality images typically absent in photo sharing platforms. Thus, in this paper, we look more deeply into user intent and build a search intent taxonomy in text-based Web image search systems. As far as we know, this work is among the first to discuss the image search intents in Web search engines.

\subsection{User behavior in image search}\label{subsection:userbehaviorinimagesearch}
Several studies analyze the user behavior logs of image search engines~\cite{andre2009designing,o2016leveraging,goodrum1999visual,pu2005comparative,tjondronegoro2009study}. Many features, such as query reformulation patterns, session length, and the number of viewed result pages are recorded and investigated. Compared to (text-based) general Web search, image search leads to shorter queries, tends to be more exploratory, and requires more interactions. Interactions with image search result pages contain abundant implicit user feedback. Previous studies on multimedia search~\cite{hua2013clickage,jain2011learning,pan2004determinants} explored user click-through data to bridge user intent gaps for image search. \citet{o2016leveraging} proposed a number of implicit relevance feedback features based on additional interactions including hover-through rate, converted-hover rate, and referral page click-through to improve image search result ranking performance. These findings help us to  understand how users search for images on the Web, but do not capture variation among types of image search intent.

Although image search tends to be more exploratory, image searches can also be intent-directed~\cite{andre2009designing}. \citet{park2015large} analyzed a large-scale query log from Yahoo Image Search to investigate user behavior toward different query types and identified important behavioral differences across them. The major difference between their work and ours is that they tried to link query type based behavior to two of four classes of image search intent proposed by \citet{lux2010classification}, which had been derived from user queries on a photo sharing platform. Since user behavior in search is heavily dependent on intent~\cite{liu2010search}, it is likely that behavior varies across different search intents. Understanding such differences, and how they relate to image search intent in Web search engines, is the main focus of this paper. 

\section{USER INTENT TAXONOMY}
\label{section:userintenttaxonomy}

We start by explaining how we used an open-coded discussion methodology to arrive at our proposed image search intent taxonomy. We then verify the proposed taxonomy through a user survey involving over 200 people and through a Web image search log analysis.

\negskip
\subsection{Establishing an intent taxonomy for image search}\label{subsection:taxonomyestablishment}
To build our intent taxonomy, we conducted both an online survey and a Web search log analysis. We applied the methodology proposed in~\cite{russell2009task} to generate our criteria to categorize image search intents. In~\cite{russell2009task}, an open-coded discussion was performed by Web research professionals based on 700 anonymized sessions aimed at identifying search task categories in general Web search. 

\negskip
\subsubsection{User survey}
\label{subsubsection:userstudy}
Besides several basic demographic questions, we ask participants to answer two open-ended questions to describe their most recent image search experience. As shown in~\cite{lux2010classification}, interviews with search users can bring us a more comprehensive understanding of search intent.
\begin{itemize}[nosep,leftmargin=12pt]
	\item Please describe your most recent image search experience with as many as details as possible (e.g., time, place, motivation).
	\item Please provide all the queries you used in this search (you can look into your search history if necessary).
\end{itemize}
In order to make our survey more accurate, we suggest participants to check their search history to help recall the experience. We spread our survey through a widely used social platform (Wechat) and paid participants about US\$0.50 if they answered the questions seriously. A total of 258 people participated in our survey; after removing noise from the answers (e.g., answers that are too short or not about text-based Web image search), 211 valid cases were kept, which are from 47.9\% female users and 52.1\% male users. The age distribution of our survey participants is shown in Table~\ref{age_distribution} together with the age distribution of Chinese internet users~\cite{cnnic201638th}. From Table~\ref{age_distribution}, we can observe that the age distribution of our survey is similar to that reported in~\citep{cnnic201638th}, except that the number of people in their 20s of our survey is much higher. 

\begin{table}[t]
\centering
\caption{Age distribution of participants in our user survey and Chinese internet users according to the 38th statistical report of China internet development~\cite{cnnic201638th}.}
\label{age_distribution}
\begin{tabular}{lm{10em}<{\centering}m{10em}<{\centering}}
\toprule
\bf Age &\bf Proportion (Survey participants) &\bf Proportion (Chinese internet users)\\ 
\midrule
$(,20)$ & 28.4\% & 23.0\% \\ 
\midrule
$[20,30)$ & 61.2\% & 30.4\% \\ 
\midrule
$[30,40)$ & 5.7\% & 24.2\% \\ 
\midrule
$[40,50)$ & 4.3\% & 13.4\% \\ 
\midrule
$[50,)$ & 0.5\% & 9.0\% \\ 
\bottomrule
\end{tabular}
\end{table}

\begin{figure*}[h]
\centering
\includegraphics[width=.8\textwidth]{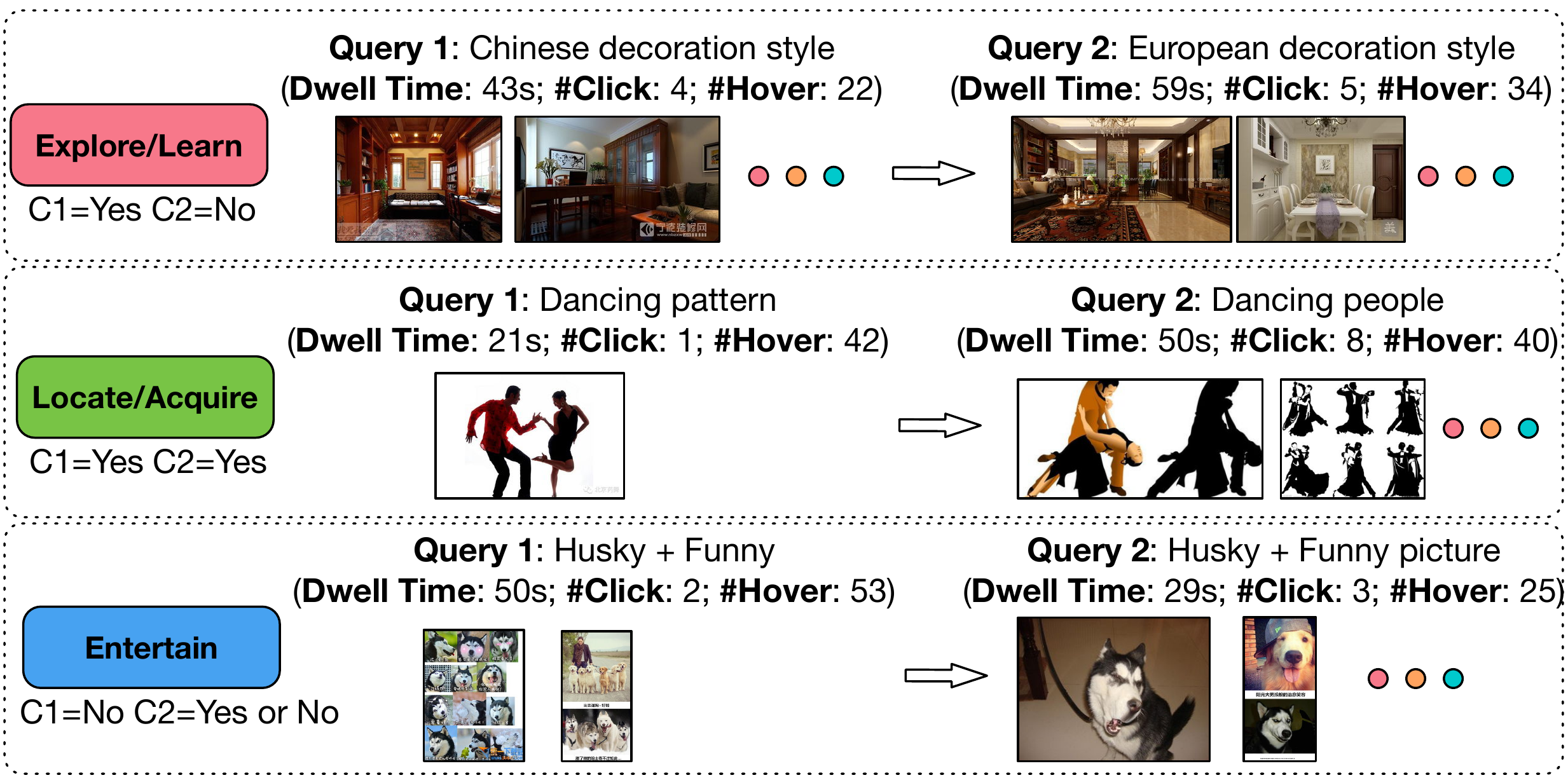}
\caption{Examples of search queries, user behaviors and clicked images in typical search sessions with different user intents (``Explore/Learn'', ``Locate/Acquire'', ``Entertain'').}
\label{taxonomy_example}
\end{figure*}

\negskip
\subsubsection{Search logs}\label{subsubsection:searchlogs}
From the search logs of a popular commercial image search engine, Sogou,\footnote{http://sogou.com} we sampled 500 search sessions during five days in April 2017. Each session contains consecutive queries produced from a single user within 30 minutes. We removed 25 pornographic search sessions and retained the other 475 sessions to avoid disturbing participants. The number of sessions examined here is similar to~\cite{russell2009task}. The minimum session length is 1 which means the user only submits one query in this session, while the maximum session length is 29. The average session length is 3.44. We show the distribution of session length in these 475 sessions in Table~\ref{session_length} (column~2). From Table \ref{session_length}, we can observe that short sessions account for a large proportion.

\begin{table}[t]
\centering
\caption{Distribution of session length of sampled search sessions after filtering pornographic searches and the Fleiss' Kappa~\cite{fleiss1971measuring} for sessions with different session length.}
\label{session_length}
\begin{tabular}{ccc} \toprule
\bf Session length&\bf Proportion&\bf Fleiss' Kappa\\ 
\midrule
1&24.4\%&0.419 \\ \midrule
2&30.1\%&0.311 \\ \midrule
3&13.3\%&0.317 \\ \midrule
4&9.1\%&0.463 \\ \midrule
5&6.9\%&0.237 \\ \midrule
$>$ 5&16.2\%&0.408 \\ \midrule
All&100\%&0.375 \\ \bottomrule
\end{tabular}
\end{table}

\subsubsection{Categorization criteria}
\label{subsubsection:categorizedcriterion}
After collecting the online survey data and search logs, a group of three web research professionals was recruited to review all 211 survey data cases and 475 search sessions. The researchers read each session closely and discussed to determine criteria to categorize user intent. Following \cite{russell2009task} we performed several iterations of this open-coded discussion. In each iteration, the proposed criteria and corresponding intent taxonomy were fine-tuned to cover as many search sessions as possible and to be easy to state and understand. Based on the results of the discussion, two criteria were proposed to categorize user intent:
\begin{description}[nosep,leftmargin=14pt]
	\item[Criterion 1] Is the user's search behavior driven by a clear objective?
	\item[Criterion 2] Does the user need to download the image for further use after the search process?
\end{description}
We first divided user intent into two groups according to Criterion~1. In some cases, people regard image search as an option for entertainment. They can freely browse the image search results without prior requirements for the images. One can enjoy photos of her/his favorite stars or some humorous images without following someone's account on social media communities, which makes image search an easy way to relax. We label this kind of user intent with ``\emph{Entertain}.'' 

In other cases, people may want to find specific images that should meet several requirements they have in mind. We further discussed intent categorization by considering Criterion 2 for such cases. In some cases, people have to download images for further use. They may already have captions for these images. For example, they may want to write a report on the 2016 US presidential elections and need to find an image that shows two presidential candidates in a television debate. We call user intent in this group ``\emph{Locate/Acquire}:'' people want to find and download images for which they already have some requirements, to complete some tasks. 

For other tasks, people are capable of satisfying their information need without downloading images. Through image search engines, they can obtain, check or compare information by examining images in result pages only. To be more specific, consider three queries ``New Ferrari 458,'' ``a flower with purple stamen'' and ``decoration style.''. The first query may come from the need that people want to know about the appearance of the new Ferrari car. When people want to find out the name of a flower that they saw and remember its major characteristics, they may use the query ``a flower with purple stamen.'' For the third one, people may want to compare different characteristics of different decoration style in order to find the best one for their houses. We call this kind of user intent ``\emph{Explore/Learn}:'' people expect to obtain informational gain from search result pages and their need can be satisfied without downloading images.

According to Criterion~1 and~2, we group user intents into three categories as follows. We use $C1$ and $C2$ to denote the answer to Criterion~1 and~2, respectively (e.g., $C1$-YES means that the user's search behavior is driven by a clear objective):
\begin{enumerate}[nosep,leftmargin=14pt]
\item \textbf{Explore/Learn}. Users want to learn something, confirm or compare information by browsing images. ($C1$-YES, $C2$-NO)

\item \textbf{Locate/Acquire}. Users want to find images for further use. They already have some requirements about these images. ($C1$-YES, $C2$-YES)

\item \textbf{Entertain}. Users want to relax and kill time by freely browsing the image search results. ($C1$-NO, $C2$-YES or NO)

\end{enumerate}
In Fig.~\ref{taxonomy_example}, we show examples of search queries, user behaviors and clicked images in typical search sessions with different user intents according to the collected Web user behavior log data.  

Our intent taxonomy is similar to the Web search task taxonomy proposed by \citet{russell2009task}. There are important differences, however. For a start, we do not have their ``Navigate'' and ``Meta'' in our intent taxonomy, because people rarely use queries leading to a site or test web sites' capabilities in image search tasks. Furthermore, ``Find-Simple'' and ``Find-Complex'' are not in our taxonomy as they are covered by ``Explore/Learn.'' \citet{russell2009task}'s
``Find-Simple'' is a scenario where an information need can be satisfied with a single query and a single result; their ``Find-Complex'' is a scenario where a user has to search for information that requires multiple searches on related topics, inspect multiple sources, and integrate information across those sources.
Finally, we rename \citet{russell2009task}'s ``Play'' to ``Entertain'' in our intent taxonomy  as ``Play'' in Russel's taxonomy focuses more on transactional needs.  

\negskip
\subsection{Verifying the intent taxonomy}\label{subsection:taxonomyverification}

To verify our proposed intent taxonomy, we asked three annotators (who are all graduate students majored in computer science and are different from the people who proposed the categorization criteria mentioned in Section~\ref{subsection:taxonomyestablishment}) to manually annotate the search scenarios collected in the survey into our three intent categories (Locate/Acquire, Explore/Learn, Entertain). We provided them with the definitions of each user intent specified in the proposed taxonomy mentioned in Section~\ref{subsection:taxonomyestablishment}. We also gave our annotators two other choices: ``Difficult to classify'' and ``Others.'' ``Others'' indicates that the user's intent cannot be fitted into any of the three intent categories and ``Difficult to classify'' means that the user's intent seems to belong to two or more classes in our proposed taxonomy. 

For 203 out of the 211 valid online survey cases our annotators obtain a majority agreement, meaning that at least two raters assign the case into the same category. The Fleiss Kappa score~\cite{fleiss1971measuring} is 0.673 among the three annotators, which constitutes a substantial agreement~\cite{landis1977measurement}. The number of cases assigned to ``Others'' by our three annotators are 1 (0.47\%), 2 (0.94\%), 5 (2.37\%), while the numbers for ``Difficult to classify'' are 3 (1.41\%), 0 (0\%), 2 (0.94\%), respectively; a closer look at those cases reveals that the descriptions of these cases are vague. Based on these annotation results, we conclude that the proposed image search intent taxonomy covers most of users' actual intents and that the taxonomy is easy to use and apply for annotators.

We employed the same annotators to annotate our search logs. Our annotators were only shown the list of queries, i.e., they were not given hits or clicks. Again, we provided them with the definitions of each user intent in the taxonomy mentioned before. The choices of ``Others'' and ``Difficult to classify'' were also given. The Fleiss Kappa score is 0.375 among our three annotators, which constitutes a fair agreement. Also, we list the Fleiss Kappa scores for different session lengths in Table~\ref{session_length}. We can observe that when the session length is around 4, the annotation agreement is highest, leading to moderate agreement. Compared with the substantial agreement reported in survey verification, it seems that by just examining queries, annotators cannot fully capture users' intents. This result echos similar observations by~\citet{russell2009task}. The result motivated us to further investigate the signals in user behavior that can be applied to distinguish search intents; see Section~\ref{section:imagesearchwithdifferentintents}.

Through the user survey involving over 200 people and the Web search log annotations, we show that our proposed taxonomy can cover a majority of user intents and the boundary between different intents is detectable for annotators. The distribution of three intents in user survey is 27\% (Explore/Learn), 66\% (Locate/Acquire) and 7\% (Entertain). In the search log the distribution is 56\% (Explore/Learn), 39\% (Locate/Acquire) and 5\% (Entertain).  

In summary, our answer to RQ1 is that user intents in image search on the web can be grouped into three classes: Explore/Learn, Locate/Acquire, and Entertain. 

\negskip
\section{IMAGE SEARCH WITH DIFFERENT INTENTS}\label{section:imagesearchwithdifferentintents}
Armed with our intent taxonomy for image search, we address RQ2, ``How does image search behavior change with user intents?'' by conducting a lab-based user study. In this user study, we pre-design a set of tasks based on the proposed user intent taxonomy.

\negskip
\subsection{User behavior dataset}\label{subsection:userbehaviordataset}

\begin{table*}[h]
\centering
\caption{Examples of user study tasks.}
\label{task_set}
\begin{tabular}{lm{20em}m{12em}m{13em}}
\toprule
\bf Category&\bf Goal&\bf Constraint&\bf Success Criteria\\ 
\midrule
Explore/Learn&Imagine you prepare to renovate a new house. You would like to compare different decoration styles (e.g., Chinese style, Simple European style).&--&Please introduce and compare the characteristics of different decoration styles in voice\\ 
\midrule
Locate/Acquire&Please change the desktop background of this computer.&The background image should have blue sky and forest.&Change the desktop background to the required image. \\ 
\midrule
Entertain&Now take a break, you can browse some posters or photos of your favorite stars.&--&-- \\ 
\bottomrule
\end{tabular}
\end{table*}

\subsubsection{User study task}\label{subsubsection:userstudytask}
We designed 12 tasks based on the results of the user survey mentioned in Section~\ref{subsubsection:userstudy}. Each category of the proposed taxonomy accounts for 4 tasks. Examples of the user study tasks are shown in Table~\ref{task_set}. 

In order to simulate a realistic image search scenario, for the ``Locate/Acquire'' tasks, we not only ask participants to complete the search part of the tasks but also ask them to use the images they find to create some multimedia productions (e.g., a slide, a poster, a computer desktop). We provided them with frequently used software to help them make the productions, with several default settings chosen by us, including the text part and background, which guarantees that the participants only need to use the image search engine to complete their tasks. Through this setup, we want to ensure that each participant faces the same task difficulty. For example, in one of the ``Locate/Acquire'' tasks, participants are asked to make a slide about Harry Potter. We pre-set the theme of the slide as ``The movie characters of Harry Potter'' and provide the names of three characters on the slide; the participants need to find posters of these three characters and coordinate different posters and the background. For the ``Explore/Learn'' tasks, we ask participants to verbally answer certain questions related to the query to ensure that the task is done seriously. For example, in one of the ``Explore/Learn'' tasks, participants are asked to find the name of a flower that has some characteristics. As the scenario assumes that participants already saw the flower, we provide an image of the flower before they search to make sure they have a mental impression. For the ``Entertain'' tasks, participants can freely browse the image search results to relax. We only pre-set the theme of the task without any further constraints. 

\negskip
\subsubsection{Data collection procedure}\label{subsubsection:datacollectionprocedure}
In the user study, each participant was asked to complete all 12 image search tasks, which were offered in a random order. Compared with collecting data from real search logs, or by browsing plugins, the laboratory user study has a smaller scale, but it does allow us to fully control the variabilities in search tasks and information needs.

We recruited 35 undergraduate students, 13 female and 22 male, via email and online social networks, to take part in our user study. The ages of participants range from 18 to 25 and their majors include engineering, humanities and social science, and arts. All participants were familiar with basic usage of Web search engines. It took about an hour and a half to complete the user study. And we paid the participants about US\$25 after they completed all the tasks seriously.

\begin{figure}[t]
\centering
\includegraphics[width=\columnwidth]{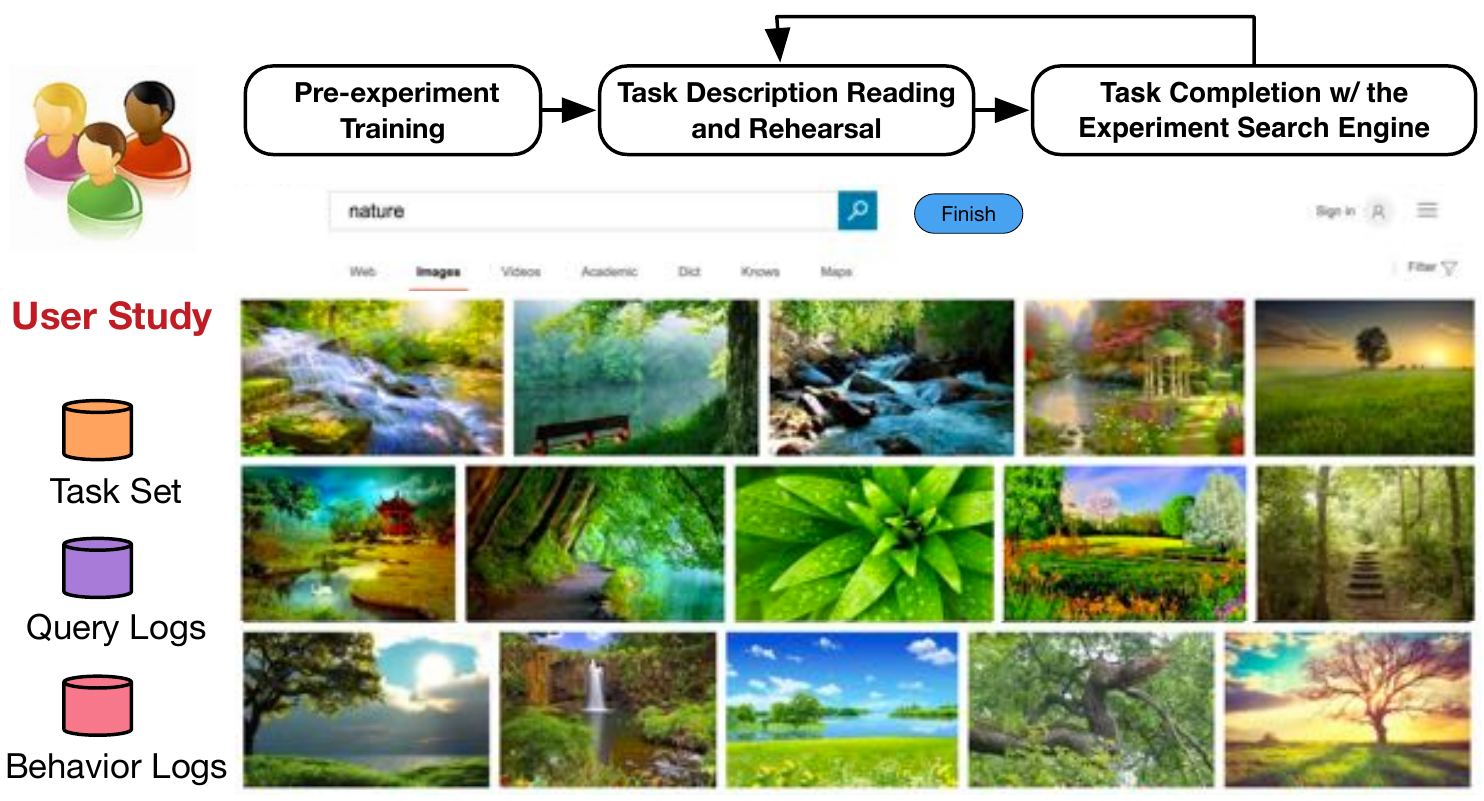}
\caption{Data collection procedure. We designed our tasks based on the proposed user intent taxonomy. With enrolled participants, we collected query logs and behavior logs.}
\label{user_study}
\end{figure}

To make sure that every participant was familiar with the experimental procedure, an example task was used for demonstration in a Pre-experiment Training stage. In the example task, we asked participants to use the image search engine to learn how to tie a tie and they were required to describe the process in voice after searching. They could browse and click the search results, and adjust the sensitivity of the mouse to the most appropriate level as well. We did not give any further instructions on which results to click on or when to end until participants were familiar with the experimental search engine. After the Pre-experiment training stage, they were asked to complete all 12 image search tasks. For each task, the participants had to go through 2 stages, as shown in Fig.~\ref{user_study}. 
Firstly, the participants should read and memorize the task description (note that the complete task description is provided to the participants). After that, they were required to describe the task in voice without viewing it. Then they can push a ``Begin Task'' button and will be redirected to an experimental search engine. Like Web image search scenario, they could scroll to move the page up and down, use the mouse to see hover text, and click a thumbnail to view and download the full-size image in the preview page. While no task time limits were imposed, they could stop searching and click the finish button when they thought that the task was completed or no further helpful information would be found. 

We injected customized JavaScript into search result pages to log mouse activities on search pages when users perform search tasks. The search system was deployed on a 17-inch LCD monitor with a resolution of $1366\times 768$. The Google Chrome browser was used to display results of search system. A database with full records of the experiments is available online for academic research.\footnote{https://tinyurl.com/y8pa8zk6}

\negskip
\subsection{User interaction features}\label{subsection:userinteractionfeatures}
From the query logs and behavior logs, we extracted 28 features that can be grouped into 6 types. We calculated all features both \textbf{globally} (i.e., using all data) and \textbf{at the early stage of a search session} (i.e., only using data captured before the first mouse scroll). Table~\ref{interaction_feature} lists the complete list of 28 features. 

\begin{table}[h]
\centering
\caption{The list of 28 features extracted from the logged implicit user interaction with the image search engine. (``*'' means that a type of feature can be extracted both in a global view and at the early stage of a search session).}
\label{interaction_feature}
\begin{tabular}{lm{14em}l} 
\toprule
\bf Feature type & \bf Description & \bf\#\\ 
\midrule
Dwell time*&The dwell time on the SERP&1 \\ 
\midrule
Mouse clicks*&Number of clicks, first click time&2 \\ 
\midrule
Mouse hover*&First and longest hover time&2 \\ 
\midrule
Mouse movement*&Min, max, mean and median of the mouse movement speed at three directions(original, X-axis, Y-axis), and the mouse movement angle and radian.&20 \\ 
\midrule
Query reformulation&Adding and deleting terms, partially change.&3 \\ \bottomrule
\end{tabular}
\end{table}

Temporal information is a widely-used feedback feature in the setting of document relevance~\citep{liu2016time,borisov-context-aware-2016}, and it varies with different search tasks~\cite{kelly2015development}. In this paper, we considered temporal information including dwell time on the SERP, time to first click, time to first hover and longest hover time on the images. We also examined the number of clicks on the SERP as clicks are strongly correlated with relevance and examination~\cite{craswell2008experimental}. Temporal information is also dependent on image search query types~\cite{park2015large}. 

Mouse movement features, which were explored in previous image search analyses~\cite{Soleymani2017Multimodal}, are investigated in this paper as well. Especially, as the placement of image search result is a two-dimensional grid instead of a linear result list, we considered not only the speed of mouse movement in the original direction but also in the X-axis (horizon) and Y-axis (vertical) direction. 

Finally, query reformulation patterns in a search session were investigated. Previous studies show that query reformulation patterns vary with search goals~\cite{huurnink2010search}, they occur frequently on image search platforms~\cite{andre2009designing}, and our data also provides evidence to support this, with approximately 76.7\% of the sessions involving more than one query. Query reformulations between consecutive queries can be grouped into four categories~\cite{jansen2007query}: adding terms, deleting terms, partial change, and complete change. Because the participants had to search for different items in some tasks, which was caused by the task description we provided, not by the user's cognitive processes, we only compare differences between adding terms, deleting terms, and partial change.

\negskip
\subsection{Statistical analysis}\label{subsection:statisticalanalysis}

\begin{figure}
\centering
\subfigure[ANOVA-$p < 0.05$]{
	\label{dt}
	\includegraphics[width=.45\columnwidth]{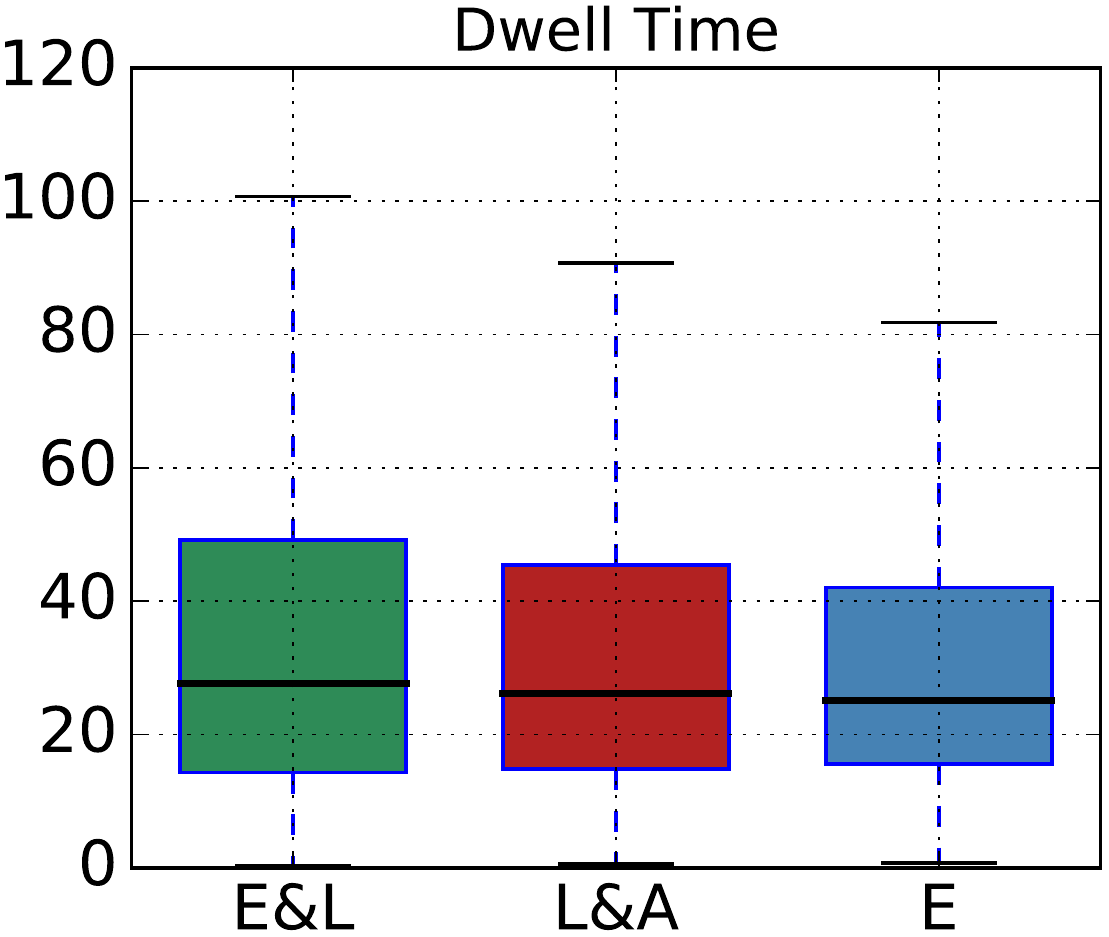}
	}
\subfigure[ANOVA-$p < 0.1$]{
    \label{dt_before}
	\includegraphics[width=.45\columnwidth]{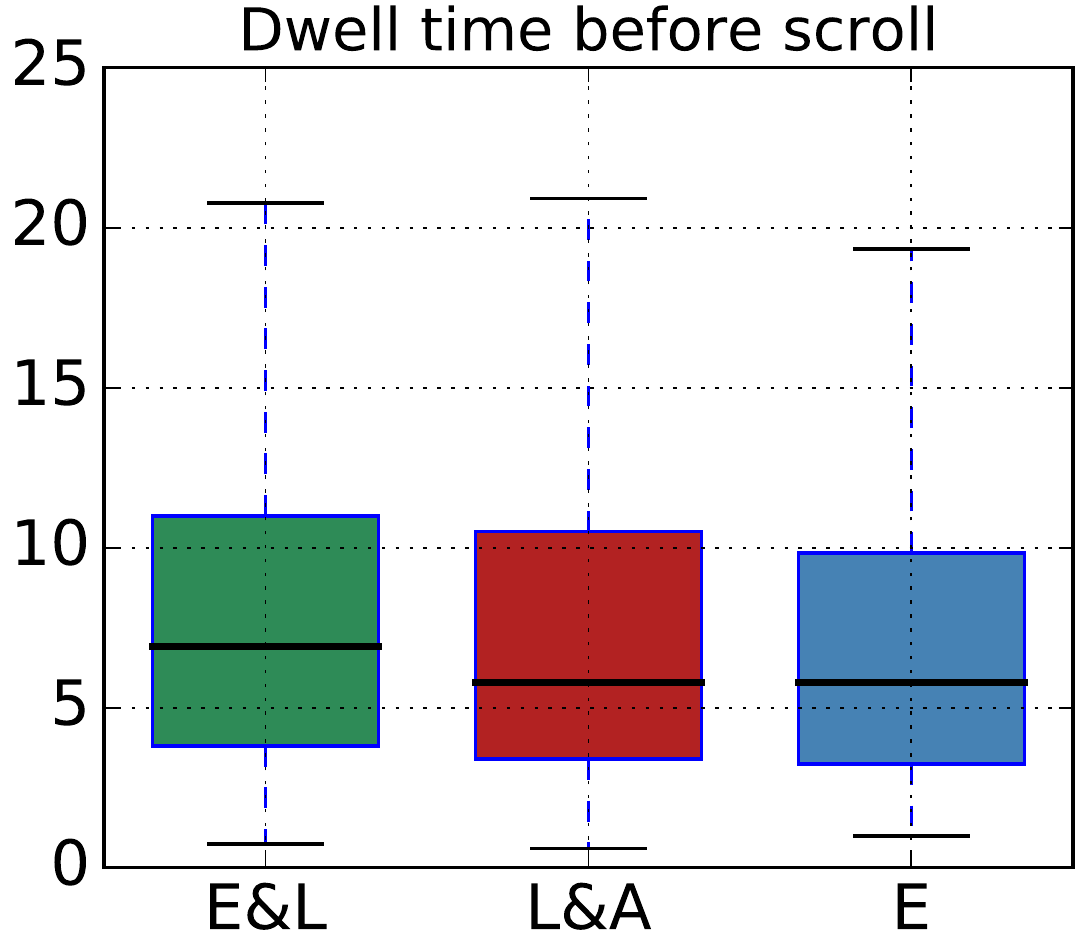}
	}
	\\[-.25\baselineskip]
\subfigure[ANOVA-$p < 0.05$]{
	\label{fc}
	\includegraphics[width=.45\columnwidth]{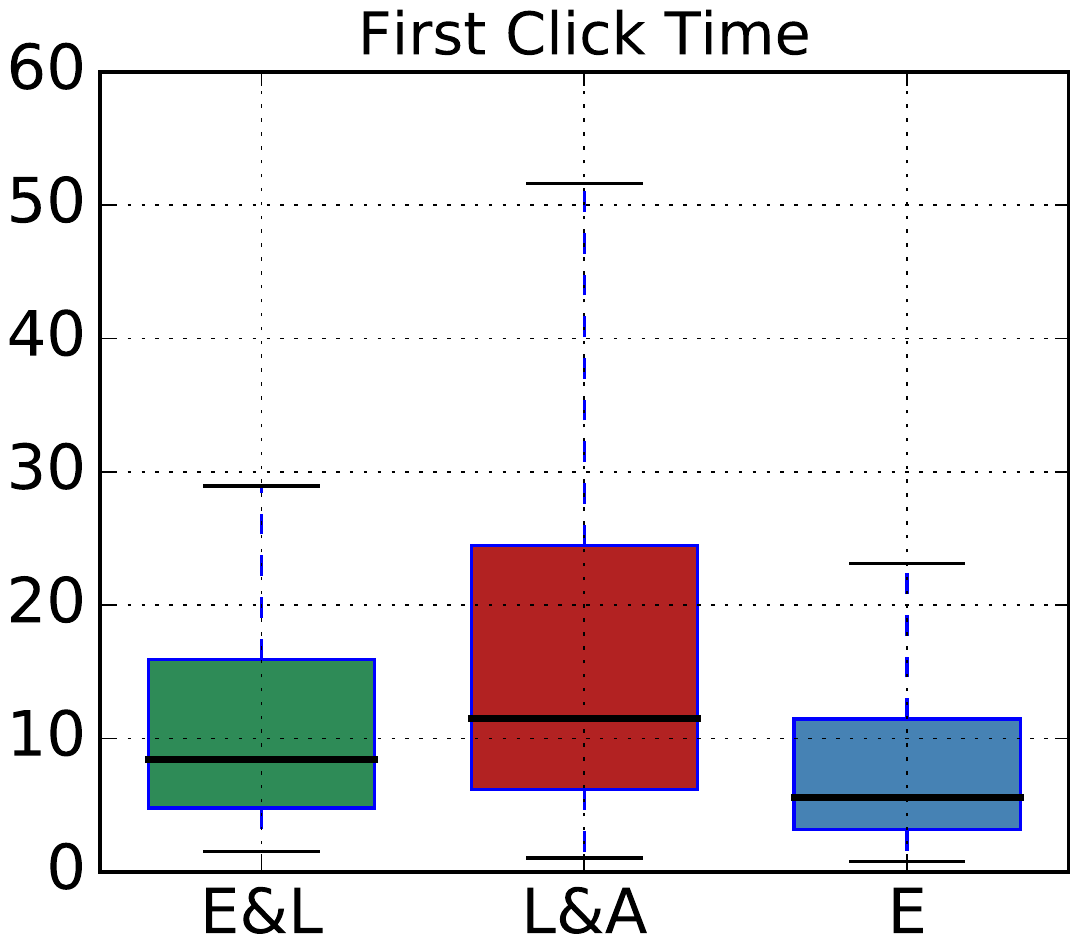}
	}
\subfigure[ANOVA-$p < 0.1$]{
    \label{fc_before}
	\includegraphics[width=.45\columnwidth]{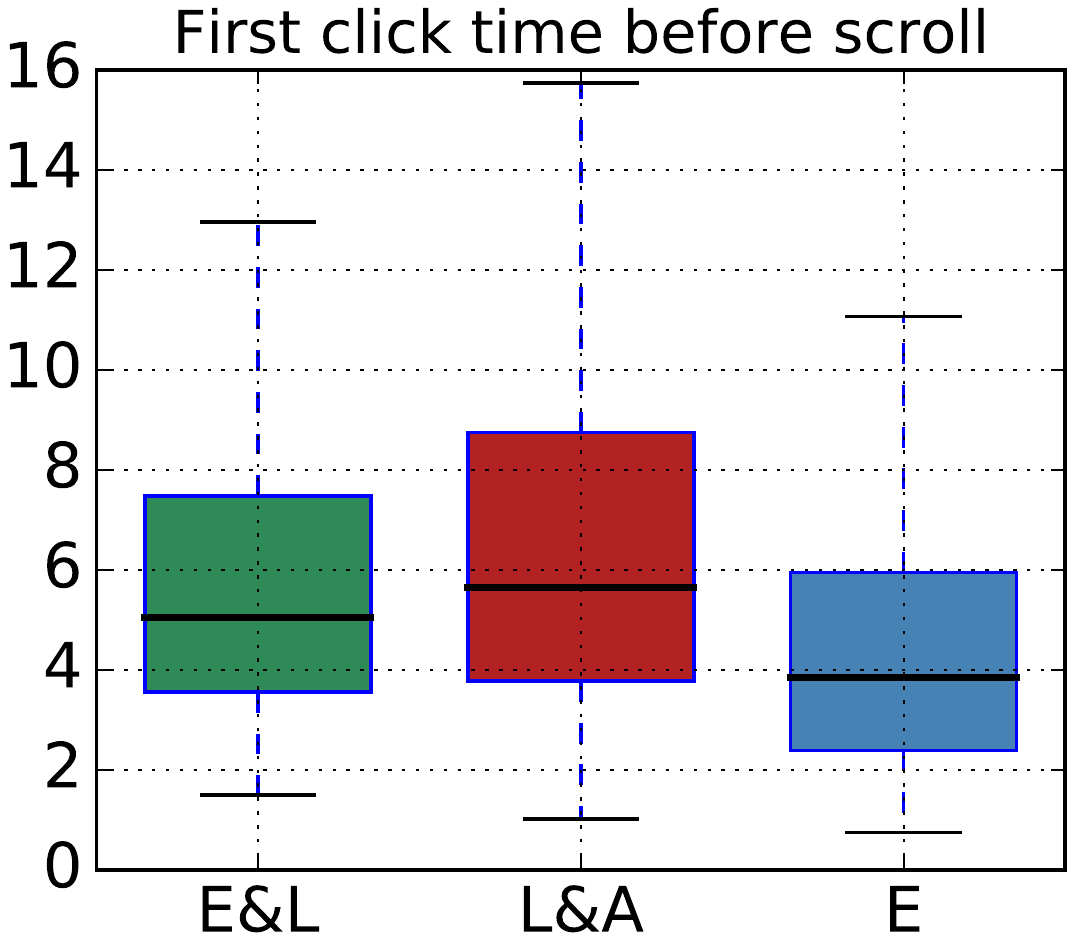}
	}
	\\[-.25\baselineskip]
\subfigure[ANOVA-$p < 0.01$]{
	\label{lh}
	\includegraphics[width=.45\columnwidth]{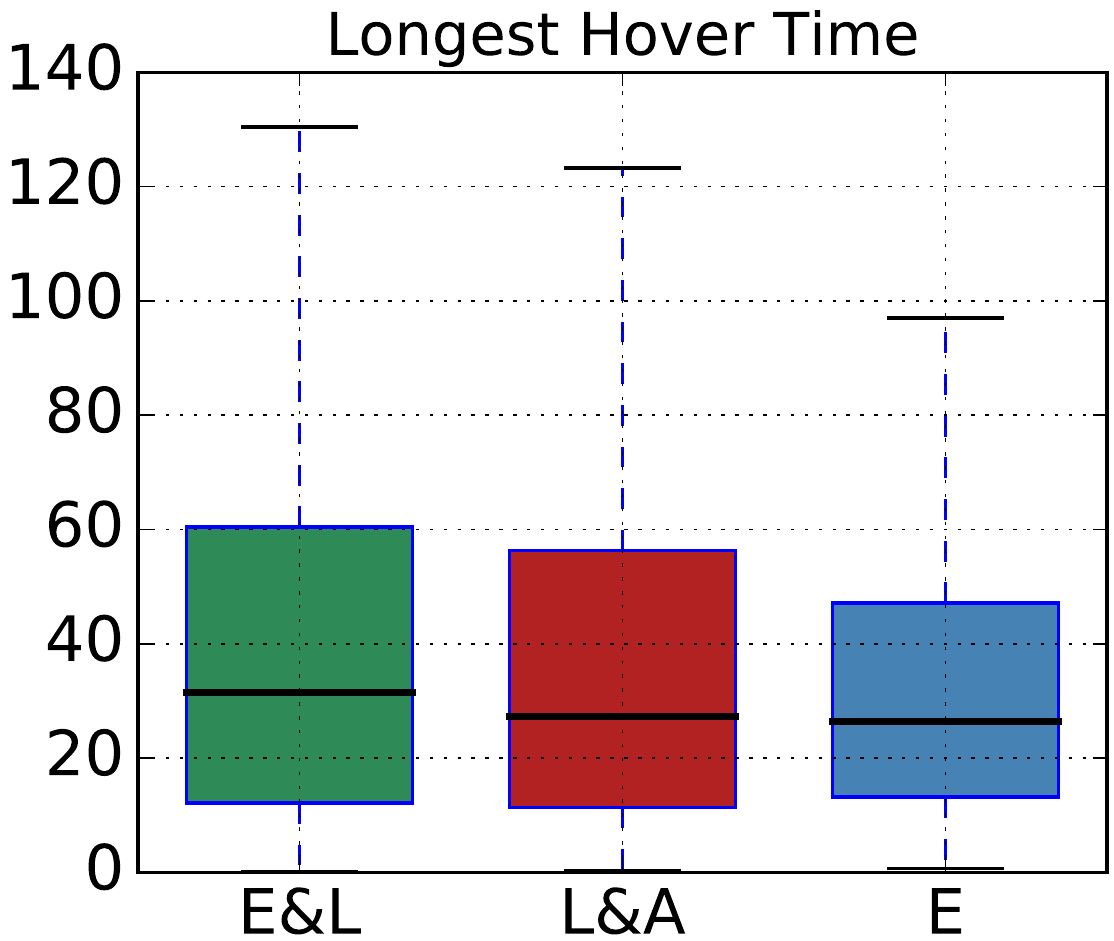}
	}
\subfigure[ANOVA-$p < 0.01$]{
    \label{lh_before}
	\includegraphics[width=.45\columnwidth]{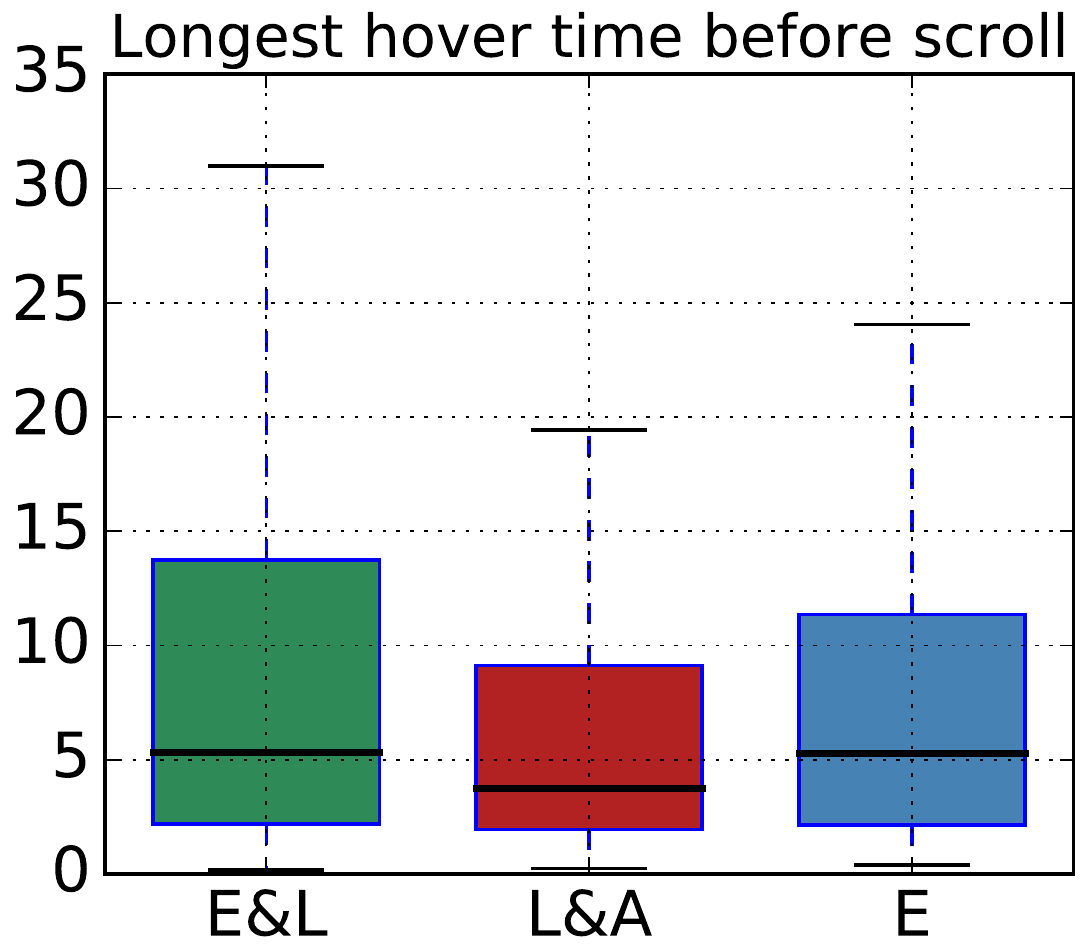}
	}
	\\[-.25\baselineskip]
\subfigure[ANOVA-$p < 0.05$]{
	\label{ms}
	\includegraphics[width=.45\columnwidth]{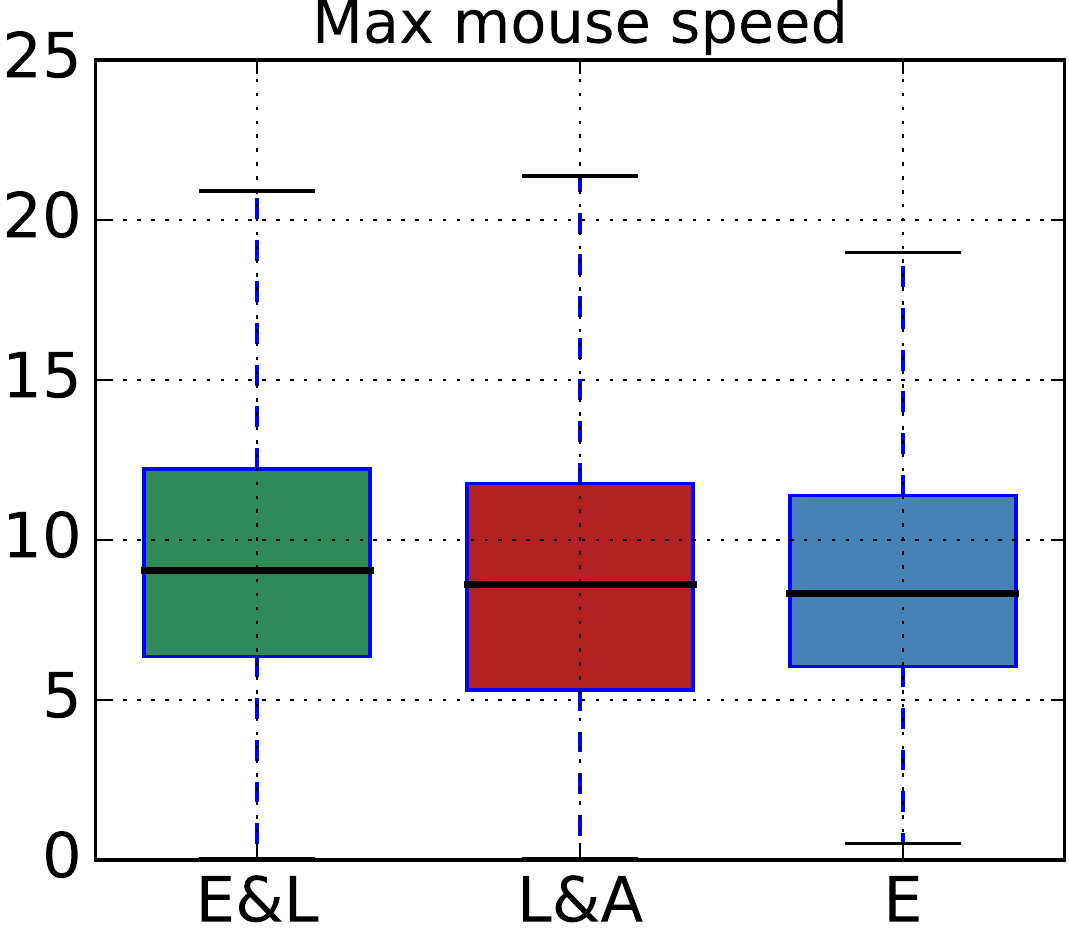}
	}
\subfigure[ANOVA-$p < 0.01$]{
    \label{ms_before}
	\includegraphics[width=.45\columnwidth]{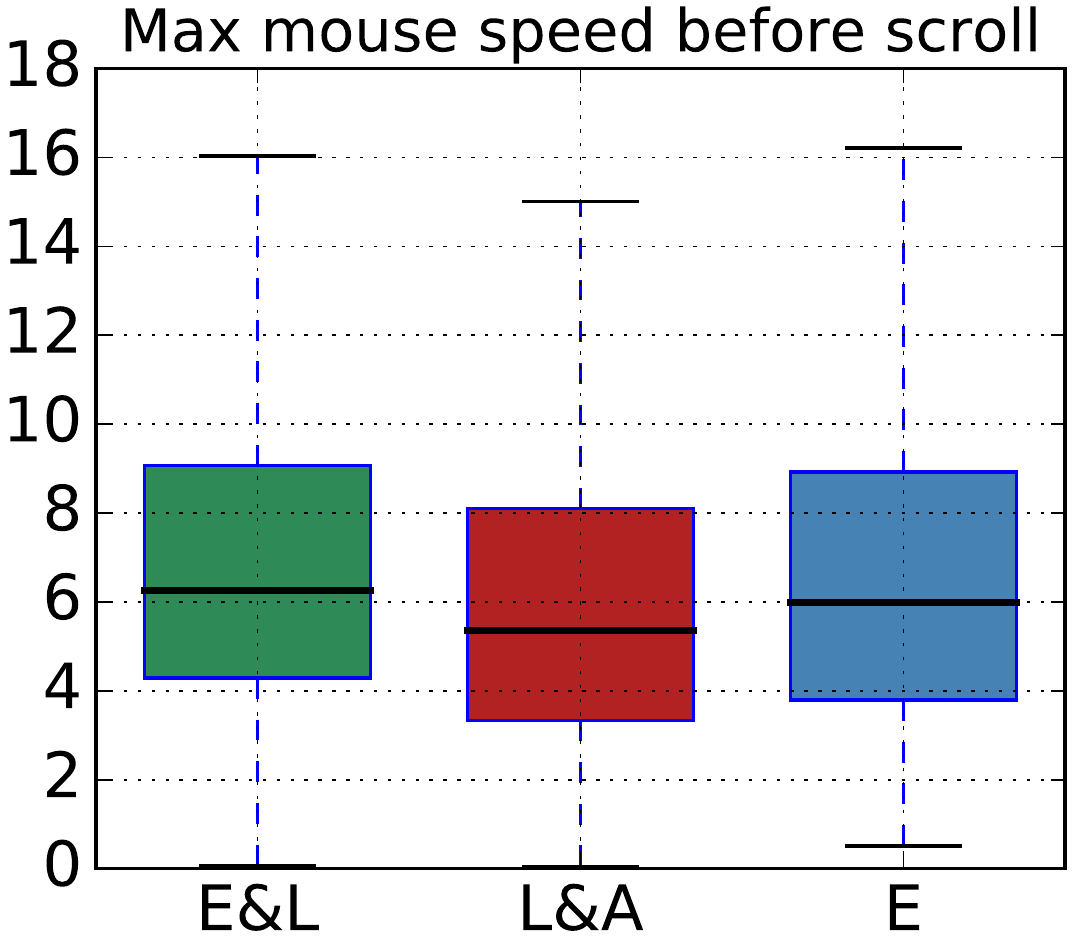}
	}
	\\[-.25\baselineskip]
\subfigure[ANOVA-$p < 0.01$]{
	\label{ma}
	\includegraphics[width=.45\columnwidth]{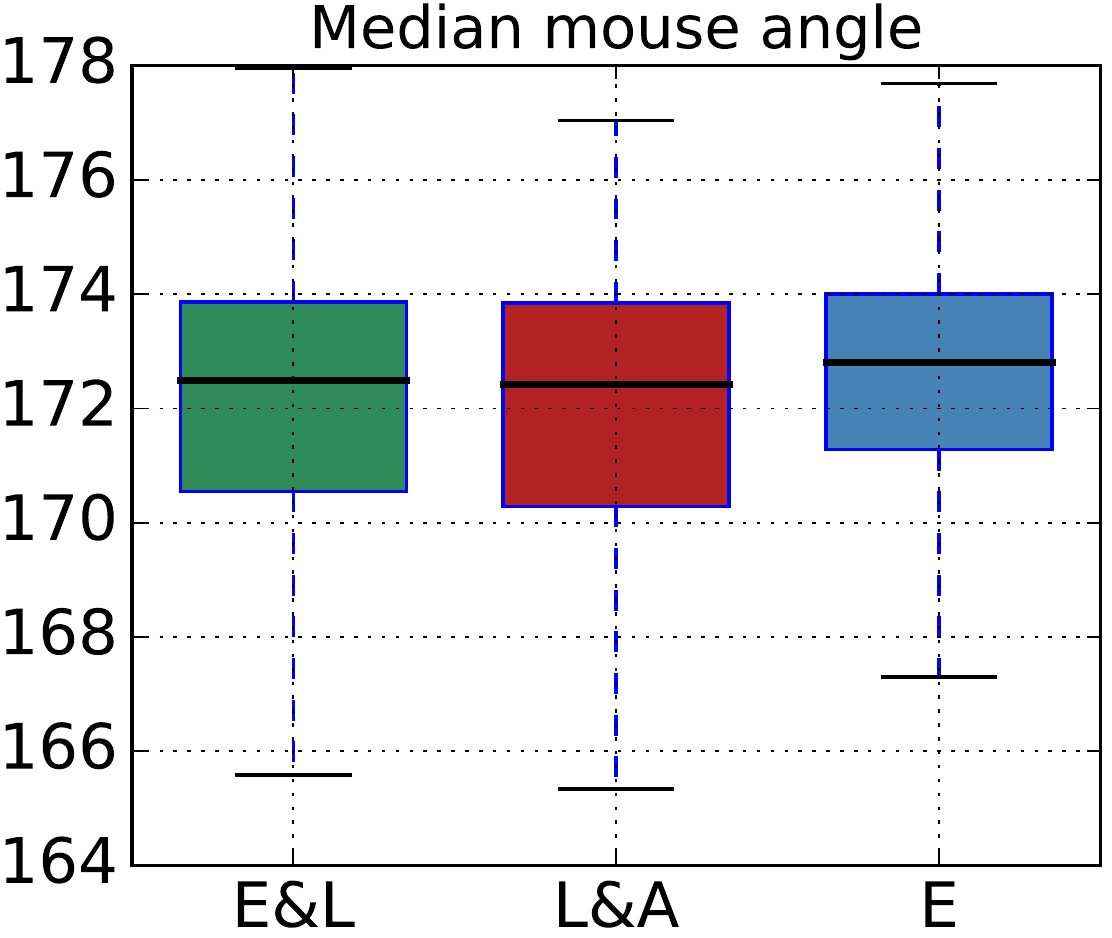}
	}
\subfigure[ANOVA-$p < 0.05$]{
    \label{ma_before}
	\includegraphics[width=.45\columnwidth]{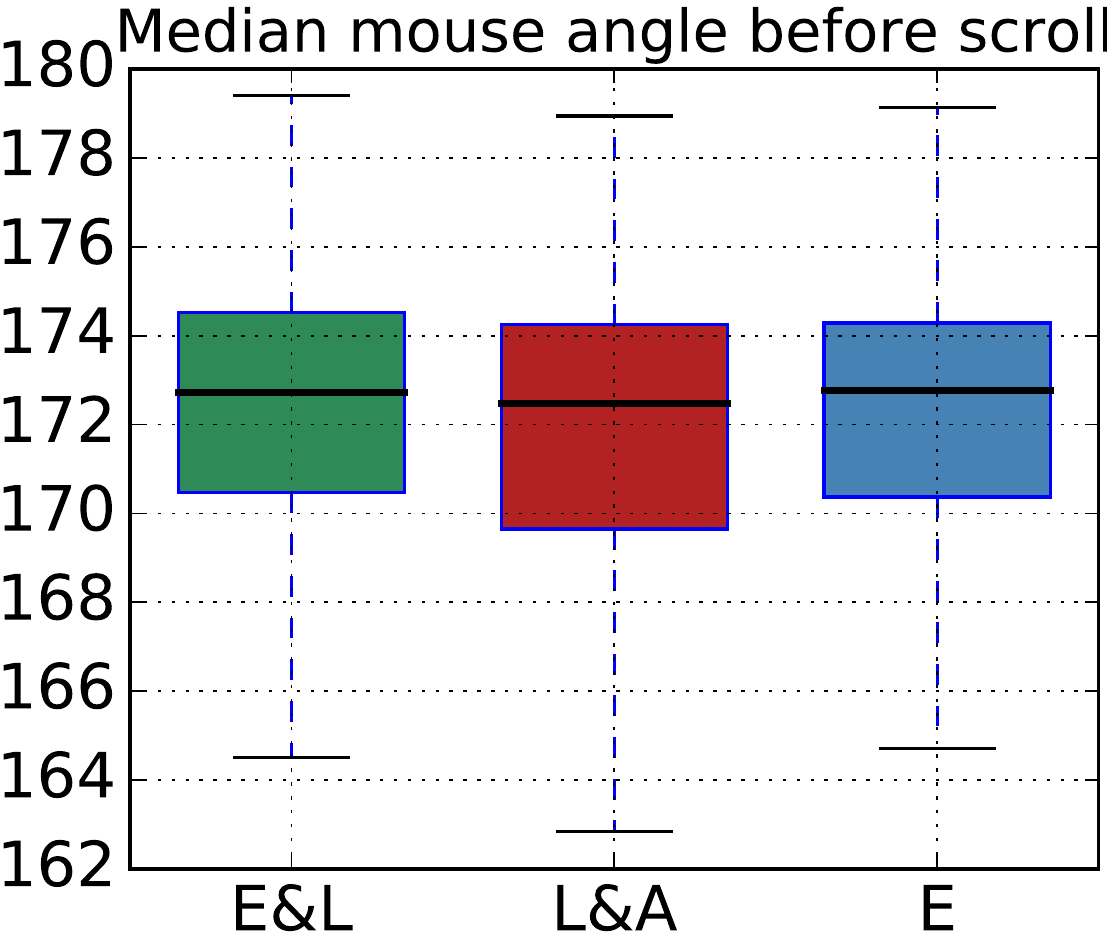}
	}
\caption{The extracted features on the SERP boxplots for queries with different search intents \emph{globally} (a, c, e, g, i) and \emph{at the early stage of search} (b, d, f, h, j). The unit on the y-axis is second for a, b, c, d, e, f; pixels per second for g, h; and number of degree for i, j. E\&L denotes Explore\&Learn, L\&A denotes Learn\&Acquire and E denotes Entertain.}
\label{box_plot}
\end{figure}

In this subsection, we report the relationship between the extracted features and search intents using box plots. We also perform a series of one-way ANOVA tests and pairwise t-tests to determine the significance. We show the box plots in Fig.~\ref{box_plot}. The difference in query reformulation patterns across different search intents is plotted in Fig.~\ref{query_reformulation_fig}. The results of the ANOVA tests (ANOVA-$p$) are reported in Fig.~\ref{box_plot} and Fig.~\ref{query_reformulation_fig}. The results of the pairwise t-tests ($p$) are discussed in Sections~\ref{subsubsection:dwelltime}--\ref{subsubsection:queryreformulation}.

\negskip
\subsubsection{Dwell time}
\label{subsubsection:dwelltime}
From Fig.~\ref{dt} and~\ref{dt_before}, we can observe that the mean dwell time in ``Explore/Learn'' tasks is longer than in ``Entertain'' tasks, both globally and at the early stage of a search session ($p < 0.02$ and $p < 0.07$, respectively). And the mean dwell time in ``Locate/Acquire'' tasks is longer than in ``Entertain'' tasks globally ($p < 0.05$). Before scrolling, dwell time on the SERP is significantly different between ``Explore/Learn'' and ``Locate/Acquire'' ($p < 0.05$). Recall our criteria for categorizing user intent: as users' search behavior in ``Explore/Learn'' and ``Locate/Acquire'' tasks is driven by a clear objective,  there will exist more confirmation and comparison of image content, which results in more time spent on search engine result pages.

\negskip
\subsubsection{Mouse clicks}\label{subsubsection:mouseclicks}
The number of clicks shows significant differences between the three intent classes ($p< 0.01$). The mean number of clicks in the three classes follows this relative order: ``Explore/Learn'' $<$ ``Locate/Acquire'' $<$ ``Entertain.'' The first time to click is also a useful implicit feedback signal in differentiating different intents. From Fig.~\ref{fc} and~\ref{fc_before}, we can observe that the average first click time of queries driven by the ``Locate/Acquire'' intent is longer than for queries with another intent, both globally and at the early stage of search (both have $p< 0.001$). The average first click time in ``Explore/Learn'' tasks is longer than in ``Entertain'' tasks before scrolling ($p < 0.001$) as well. When performing a ``Locate/Acquire'' or ``Explore/Learn'' task, users already have some specific requirements about the images. Image search results are self-contained, so that users do not need to click the document as in general Web search to view the landing page. Instead, they can observe several images before deciding which ones to download or to click to see a larger version. For this reason, users will spend more time on the search result page before the first click. 
\negskip
\subsubsection{Mouse hover}\label{subsubsection:mousehover}
Unlike the time to first click, time to first hover shows no significant differences between the three intents. However, the mean longest hover time in ``Entertain'' tasks is shorter than in ``Locate/Acquire'' and ``Explore/Learn'' tasks, both globally ($p < 0.001$) and at the early stage of search ($p< 0.01$), as shown in Fig.~\ref{lh} and~\ref{lh_before}, respectively. At the early stage of a search session, the mean longest hover time in ``Locate/Acquire'' is shorter than in ``Explore/Learn'' tasks ($p < 0.001$). As hovering on a document can be regarded as a signal that users are inspecting it; our results may be caused because ``Explore/Learn'' tasks require more complex cognitive processes, hence users may need to compare the image content with the background knowledge and mental impressions of their task.

\negskip
\subsubsection{Mouse movement}\label{subsubsection:mousemovement}
The speed of mouse movement varies between intents. As shown in Fig.~\ref{ms} and~\ref{ms_before}, the average maximum speed of mouse movement in ``Entertain'' tasks is lower than in ``Explore/Learn'' tasks ($p < 0.02$) and ``Locate/Acquire'' tasks ($p < 0.08$), globally. And at the early stage of a search session, ``Explore/Learn'' tasks obtain a higher average max speed than ``Locate/Acquire'' tasks ($p < 0.001$). Besides in the original direction, the speed of mouse movements also shows significant differences along the X-axis and Y-axis (e.g., ``Explore/Learn'' tasks receive the highest average mean speed of mouse movement along the Y-axis while ``Locate/Acquire'' tasks receive the lowest, both globally ($p < 0.001$) and before scrolling ($p < 0.05$)). For the angle of the mouse movement, the median angle of mouse movement between the three intents is significantly different ($p< 0.01$, $p< 0.05$, and $p< 0.01$, respectively), globally. And the differences are also significant between ``Locate/Acquire'' and ``Entertain'' ($p < 0.05$), before scrolling. Thus, mouse movement patterns have the potential to help us recognizing image search intents.

\negskip
\subsubsection{Query reformulation}\label{subsubsection:queryreformulation}
\begin{figure}[h]
\centering
\includegraphics[width=\columnwidth]{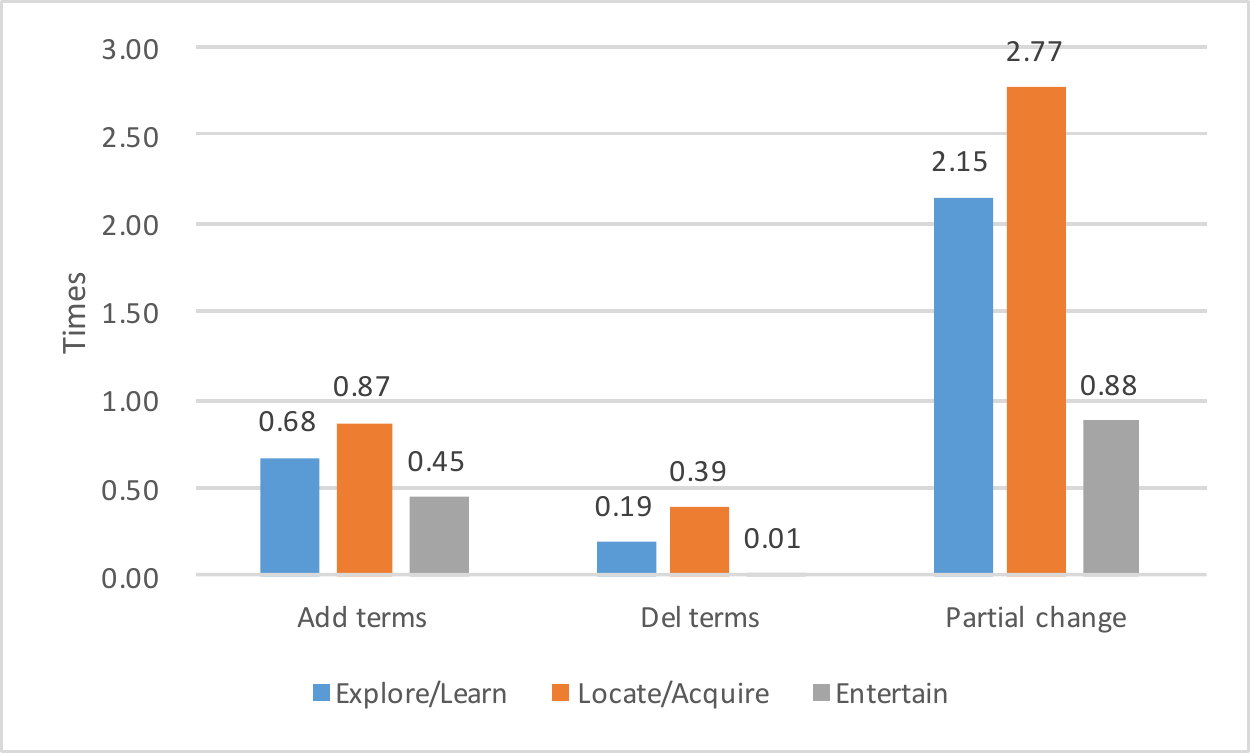}
\caption{Average number of times of query reformulations per task, across different search intents (ANOVA-$p < 0.01$ for every type of query reformulation).}
\label{query_reformulation_fig}
\end{figure}

Fig.~\ref{query_reformulation_fig} shows the average number of times a query reformulation occurs, across different search intents. The numbers indicate the number of reformulations per task, on average (globally). We observe that for ``Locate/Acquire'' tasks, participants tend to refine queries more frequently, which serve as evidence of focused search behavior. In ``Locate/Acquire'' tasks, users need to find the most appropriate images to create some productions. For example, images used as materials in designing a poster should fit the background and other materials, which means that the user needs to try different styles of images until the poster looks aesthetically acceptable. Thus, ``Locate/Acquire'' tasks receive a larger number times for query reformulation. In contrast, in ``Entertain'' tasks, the users' search behavior is not driven by a clear objective. This more exploratory, browsing-like behavior results in a smaller number of query reformulations. We performed a paired two-tailed t-test to verify the significance of the observed differences: $p<0.01$ for all comparisons except for for differences in adding terms and deleting terms between Explore/Learn and Locate/Acquire ($p < 0.05$).

In summary, through our user study, we collected query logs and behavior logs. Based on these data, we are able to answer $RQ 2$ by investigating several frequently used implicit signals in general Web search. We find that temporal information and mouse movement patterns are useful in distinguishing search intents. Also, the cognitive process under different intents may result in different types and numbers of times of query reformulation. 

\negskip
\section{IMAGE SEARCH INTENT PREDICTION using Early Stage Features}\label{section:imagesearchintentprediction}
The features discussed in Section~\ref{subsection:statisticalanalysis} show potential in helping us identifying search intent automatically. In this paper, we utilize interaction features at the early stage of a search session to build an early stage user intent recognition system aimed at addressing $RQ 3$. We aim to predict intent at the query level. If our features are effective, this system can be practical for an image search engine to rerank its results even before users begin to scrolling, which will likely improve the satisfaction of users. 

We compare different combinations of features. Due to a lack of space we do not consider all features described in Table~\ref{interaction_feature} but some natural groupings only. In particular, we use ``Time'' to denote the combination of dwell time, time to first click and time to first hover. And as the number of clicks and first hover time show no significant difference under different intents at the early stage of a search session, we do not include it in the set of features that we consider. Also, we do not use the query reformulation patterns because we want to predict the user intents at the query level. We concatenate various features into a long feature vector to fuse all features (which is known as ``early fusion" of different feature groups~\cite{Soleymani2017Multimodal}). As this task can be treated as a multi-class classification problem, we apply a gradient boosting classifier~\cite{mason2000boosting} and perform 10-fold cross validation. We assign the label from the majority class to all the instances to generate a baseline. 

The results are shown in Table~\ref{classification_performance}. We can observe that our recognition system with user behavior features outperforms the baseline significantly ($p< 0.001$). And temporal features are more effective than the other two combinations. Fusing all features together achieves better prediction results than only using a single feature group. However, early fusion does not lead to a large increase in performance over the best single group of features. We compared another fusion method, i.e., ``late fusion" in which each feature group has its own classifier and the output of all classifiers are combined to obtain a final result. We used the weighted sum of scores for ``late fusion," similar to \cite{Soleymani2017Multimodal}. The results are also shown in Table~\ref{classification_performance}. We see that ``late fusion" and ``early fusion'' perform very similarly. It is worth pointing out that in absolute terms our performance figures are similar to other intent classification tasks considered in the literature, such as \citep{cheng2017predicting,Soleymani2017Multimodal}, even though we use a much sparser signal than~\citep{cheng2017predicting} and, unlike \citep{Soleymani2017Multimodal}, only use features that can be collected in real-world scenarios.

Finally, concerning  $RQ 3$, we have found that based on interactive features at the early stages of a search session, we can recover user intents effectively through a combination of temporal features and mouse movement features.

\begin{table}
\centering
\caption{Classification performance based on the user interactions features in terms of weighted average of F-1 score. Best results are in boldface. (E\&L denotes Explore\&Learn, L\&A denotes Learn\&Acquire and E denotes Entertain).}
\label{classification_performance}
\begin{tabular}{m{5em}m{3em}<{\centering}m{5em}<{\centering}m{4em}<{\centering}m{4em}<{\centering}} 
\toprule
\bf Features&\bf All classes& \bf E\&L vs.\ L\&A & \bf E\&L vs.\ E & \bf L\&A vs.\ E \\ 
\midrule
Baseline&0.28&0.44&0.42&0.54 \\ 
\midrule
Time (\#3)&0.42&0.56&0.56&0.64 \\ 
\midrule
Mouse move speed (\#12)&0.40&0.55&0.54&0.60 \\ 
\midrule
Mouse move angle (\#8)&0.38&0.55&0.54&0.61 \\ 
\midrule
Late fusion all features&0.44&0.57&0.56&0.64 \\ \midrule
Early fusion all features&\textbf{0.45}&\textbf{0.58}&\textbf{0.58}&\textbf{0.65} \\ \bottomrule
\end{tabular}
\end{table}

\negskip
\section{CONCLUSION AND FUTURE WORK}\label{section:conclusionandfuturework}
In this paper, we proposed a new user intent taxonomy for image search and verified the taxonomy through a user survey involving over 200 people. Based on a lab-based user study, we discovered significant differences in user behavior under different intents. Finally, we used these behavioral signals to recover user intents and achieved promising results.

\negskip
\subsubsection*{Implications.} As user intents can be different in image search scenarios, considering an evaluation metric appropriate for different intents can be beneficial. Also, recommender systems could prioritize showing specific, targeted content to users based on their search goals. Last but not the least, the optimization goal of search engines should be designed according to different search intents. For example, for users with an ``Entertain'' intent, the goal may be to keep them engaged with the image search engines for as long as possible. This is where our work contributes.

\negskip
\subsubsection*{Future work.} Interesting directions for future work include investigating the intent prediction beyond the early stage, e.g., by incorporating content features of query and images besides user interaction signals. Moreover, we plan to design strategies to rerank image search results according to different intents, with the aim to improve user satisfaction. Also, a parallel comparison between video search intents and image search intents might be interesting as both belong to multimedia search.
\end{spacing}

\medskip
\begin{spacing}{0.96}
\noindent\small
\textbf{Acknowledgments.}
This work is supported by the Natural Science Foundation of China (Grant No. 61622208, 61732008, 61532011), National Key Basic Research Program (2015CB358700), 
Ahold Delhaize,
Amsterdam Data Science,
the Bloomberg Research Grant program,
the Criteo Faculty Research Award program,
Elsevier,
the European Community's Seventh Framework Programme (FP7/2007-2013) under
grant agreement nr 312827 (VOX-Pol),
the Microsoft Research Ph.D.\ program,
the Netherlands Institute for Sound and Vision,
the Netherlands Organisation for Scientific Research (NWO) under project nrs.\ 
13675, 
612.\-001.\-116, 
CI-14-25, 
652.\-002.\-001, 
612.\-001.\-551, 
652.\-001.\-003, 
and
Yandex.
All content represents the opinion of the authors, which is not necessarily shared or endorsed by their respective employers and/or sponsors.
\end{spacing}

\bibliographystyle{ACM-Reference-Format}
\bibliography{WSDM2018} 

\end{document}